\documentclass[doublecol]{epl2} 
\usepackage{subfigure}
\usepackage{inputenc}
\usepackage{url}
\usepackage{amsmath,amssymb}
\title{Unpredictability and basin entropy}

\author{Alvar Daza\inst{1,2} \and Alexandre Wagemakers\inst{1} \and Miguel A.F. Sanju\'{a}n\inst{1,3}}

\institute{ 
 \inst{1} Nonlinear Dynamics, Chaos and Complex Systems Group, Departamento de F\'{i}sica, Universidad Rey Juan Carlos, M\'{o}stoles, Madrid, Tulip\'{a}n s/n, 28933, Spain. \\
\inst{2} Department of Physics, Harvard University, Cambridge, Massachusetts 02138, USA.\\
\inst{3} Department of Applied Informatics, Kaunas University of Technology, Studentu 50-415, Kaunas LT-51368, Lithuania
}

\abstract{
The basin entropy is a simple idea that aims to measure the the final state unpredictability of multistable systems. Since 2016, the basin entropy has been widely used in different contexts of physics, from cold atoms to galactic dynamics. Furthermore, it has provided a natural framework to study basins of attraction in nonlinear dynamics and new criteria for the detection of fractal boundaries. In this article, we describe the concept as well as fundamental applications. In addition, we provide our perspective on the future challenges of applying the basin entropy idea to understanding complex systems.
}

\begin{document}

\maketitle

\section{Why basin entropy?}

The idea of uncertainty has pervaded physics in the past century, as well as other scientific disciplines. Among them, we could mention climate dynamics, economy and epidemics~\cite{palmer_2022}. From the Heisenberg uncertainty principle, uncertainty is intrinsic in quantum mechanics. Some qualify it as ontological uncertainty. However, the idea of uncertainty has also revealed itself in other areas of physics, especially through the advent of chaos theory. This notion of classical uncertainty was already pointed out in its day by some Physics Nobel Prize laureates such as Richard Feynman~\cite{feynman_1963} or Max Born~\cite{born_1954,born_1955}. There are several possible sources of uncertainty in the context of nonlinear dynamical systems. The notion of sensitivity to initial conditions that has come to be understood as one of hallmarks of chaos is one of them. The existence of fractal structures in phase space where we analyze the dynamical systems is another.

Precisely, one very relevant source of uncertainty in nonlinear dynamical systems is provided by the fractal structures appearing in basins of attraction in phase space. A basin of attraction is defined as the set of initial conditions whose trajectories go to a specific attractor.  A simple and natural analogy of a basin comes from hydrology, and the notion of a river basin. A map of the river basins of a given country help us to visualize the regions where the waters go to a given river that might be understood as the attractor. In multistable systems, trajectories may have different fates due to a small perturbation or uncertainty in the initial conditions. When we have several attractors in a given region of phase space, then we have several basins that are separated by the corresponding boundaries. These boundaries can be classified as smooth basins and fractal basins, depending on the geometrical nature of the boundaries~\cite{aguirre_fractal_2009}. An important consequence of the presence of fractal basin boundaries is the unpredictability and uncertainty in the evolution of trajectories of the dynamical system. The term fractal basins is commonly used to refer to the boundaries between basins when they are fractal. Needless to say, basins of attraction are defined for dissipative dynamical systems where attractors do exist. Hamiltonian systems do not have attractors. However, trajectories in open Hamiltonian systems may have the possibility to escape the action of the potential. In these cases a similar notion of escape basins is considered as the set of initial conditions for which trajectories may escape by a certain exit. 

Now, a fundamental question arises when we try to compare a couple of basins: to ascertain which basin is more unpredictable. However, until the appearance of the novel concept of basin entropy in 2016~\cite{daza2016basin}, there was no quantitative way to identify when a given basin was more unpredictable than another. In spite of that, there are many categories to label the basins: smooth, fractal, Wada, riddled, intermingled (for a review on the topic see~\cite{aguirre_fractal_2009}). Furthermore, there were also several measures to quantify different aspects associated to the unpredictability of the basins such as the uncertainty exponent~\cite{grebogi_final_1983}, the basin stability~\cite{menck_how_2013}, the lacunarity~\cite{mandelbrot1995measures}, etc. The idea of basin entropy supposes a quantitative measure of the unpredictability of basins. The numerical value of the basin entropy basically depends on three key ingredients which are related to the boundary size, the uncertainty dimension of the basin boundaries and the total number of attractors in the specific region in phase space.

The possible applications of the basin entropy are manifold. It has been applied to numerous problems in physics~\cite{daza2018basin}, such as chaotic scattering associated to experiments of cold-atoms~\cite{daza2017chaotic}, chaotic dynamics in relativistic chaotic scattering~\cite{bernal2018uncertainty,bernal2020influence}, in astrophysics to measure the transition between nonhyperbolic and hyperbolic regimes in open Hamiltonian systems~\cite{nieto2020measuring} just to mention a few. Moreover, it could be appropriate for other problems such as the analysis of Wada structures associated to the dynamics of photons in binary black hole shadows constituting a problem of chaos in general relativity~\cite{daza2018bh} or even in dynamical systems with delay~\cite{daza2017delay}. 

\begin{figure*}
%\onefigure{HH_EPL.eps}
\subfigure[$E=0.2$]{\includegraphics[width=0.33\textwidth]{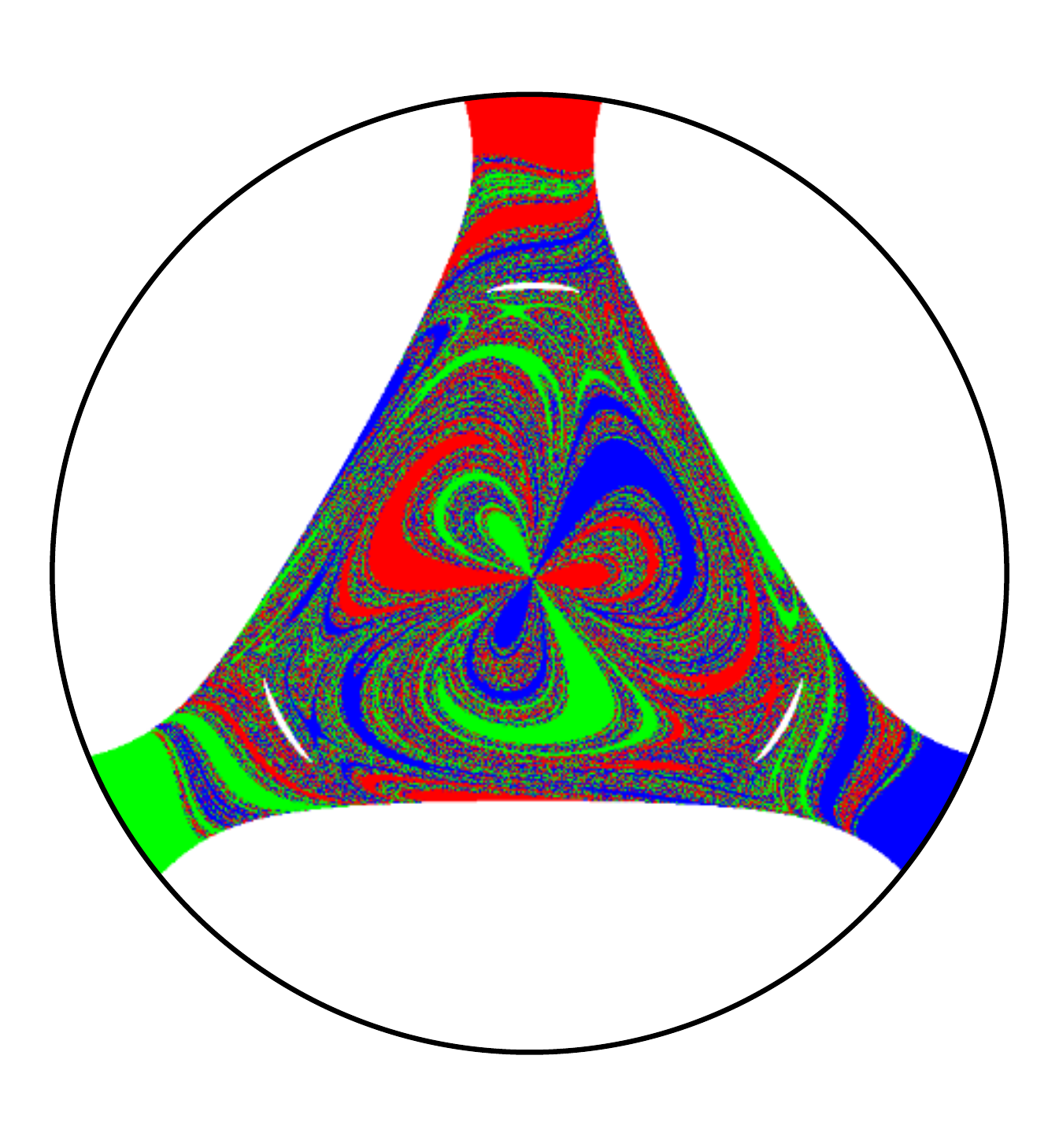}}
\subfigure[$E=0.25$]{\includegraphics[width=0.33\textwidth]{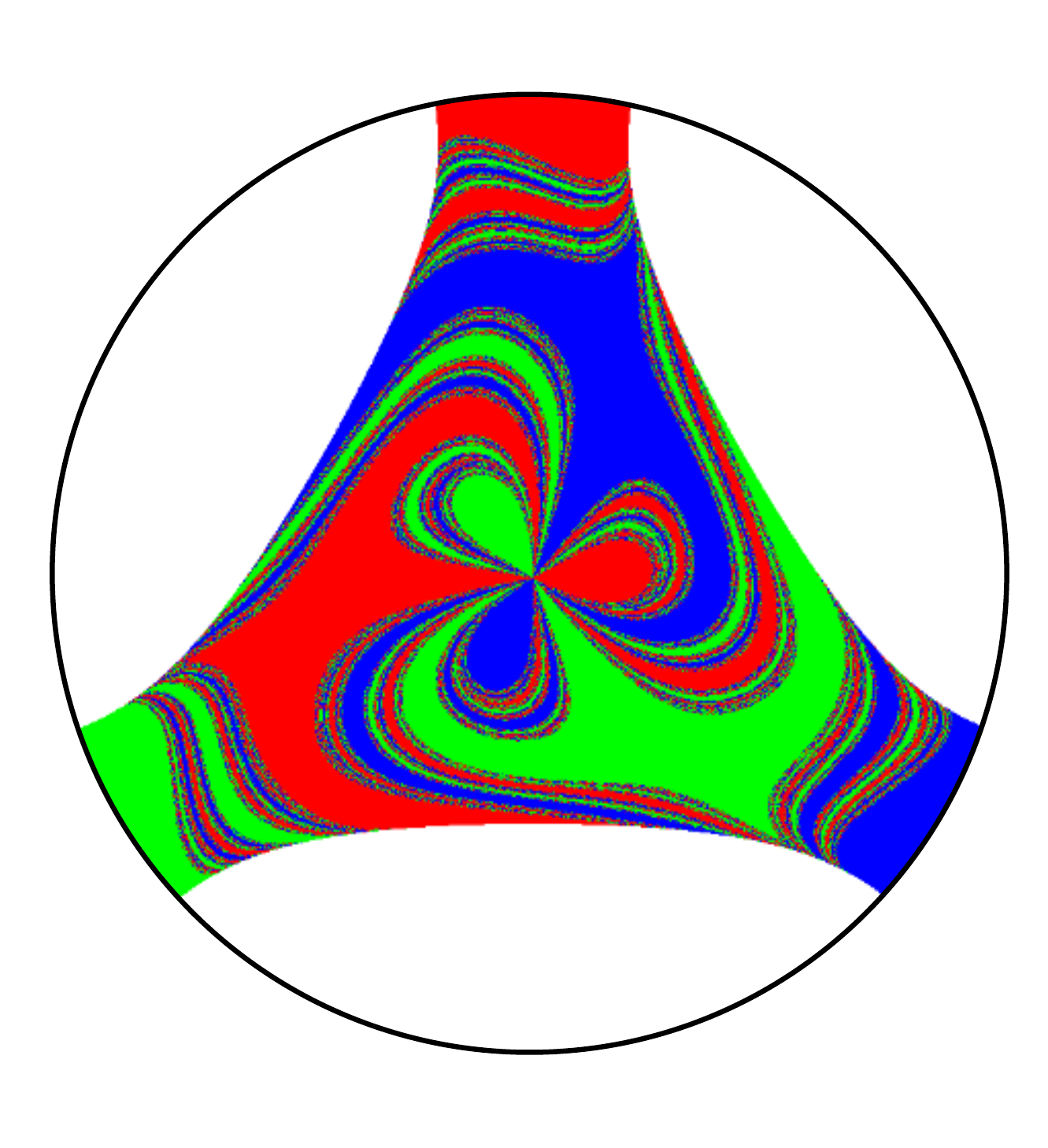}}
\subfigure[$E=0.25$]{\includegraphics[width=0.33\textwidth]{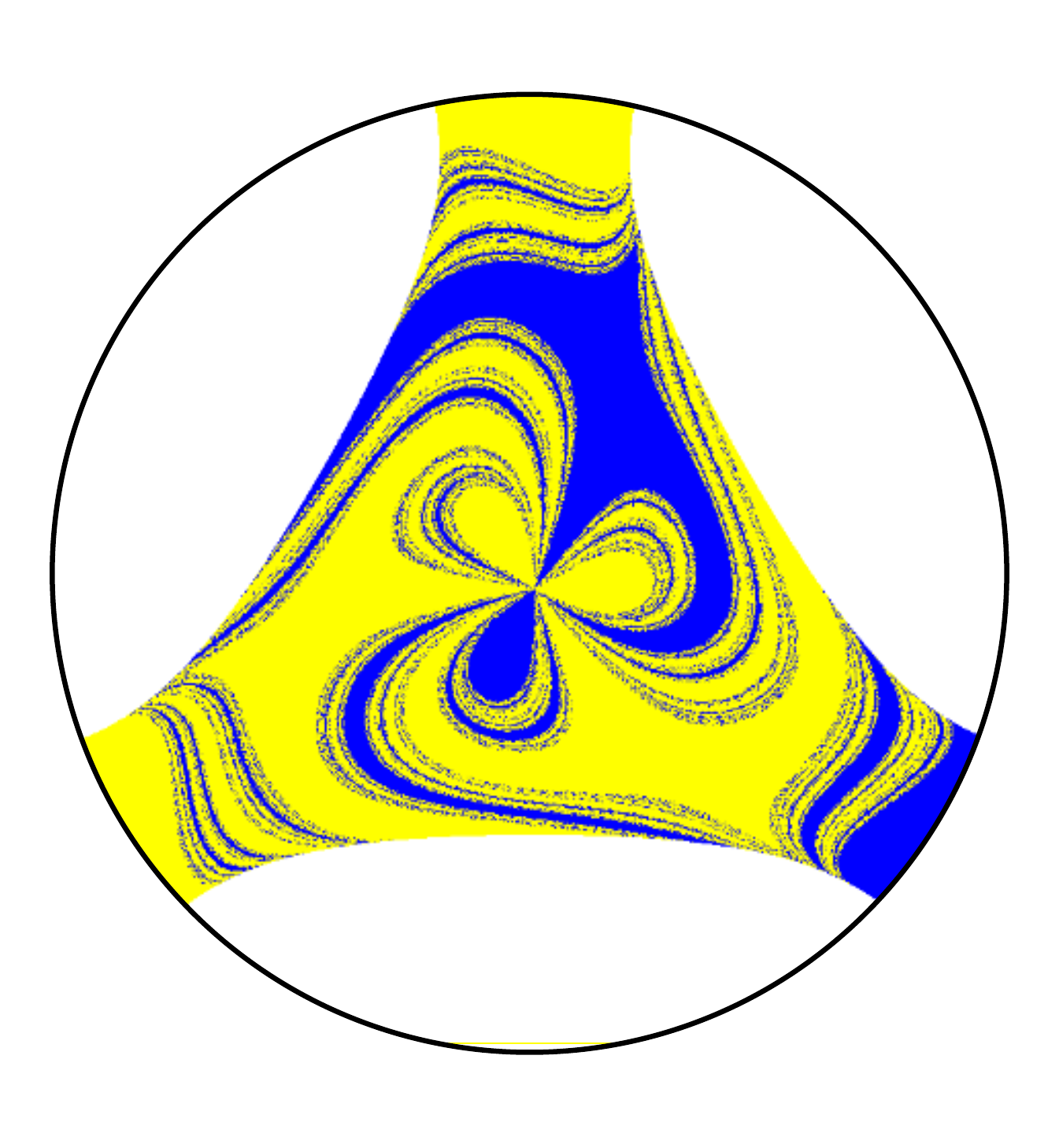}}
\caption{Three escape basins for the Hénon-Heiles Hamiltonian. (a) For $E=0.2$, which is above the escape energy $E=1/6$, the basin is highly fractalized and thus it is difficult to know which exit (red, green or blue) some initial conditions will take. (b) For a higher energy $E=0.25$, the basin is less fractalized, although the proportion of the three colors remains unaltered and equal to $\frac{1}{3}$ due to the intrinsic symmetry of the system. (c) This is the same picture as (b), but with the red and the green basins merged into yellow (for instance, one could argue that both exits are connected). Because of the merging property of the Wada basins~\cite{daza2018ascertaining}, the fractal dimension of the boundary in (c) is exactly the same as in (b). Although the area of each basin in (a) and (b) is the same, and the fractal dimension of (b) and (c) also coincides, the basin entropy correctly grasps the intuitive notion of the uncertainty associated to the three basins and classifies them $S_b(a)>S_b(b)>S_b(c)$. The computer code for reproducing the plots is available at \cite{github_code}.}
\label{fig:HHbasins}
\end{figure*}

As an example to illustrate how the basin entropy is useful to capture the intuitive idea of unpredictability associated to the basins, we can focus our attention on the escape or exit basins of the Hénon-Heiles Hamiltonian displayed in Fig.~\ref{fig:HHbasins}. 

The Hénon-Heiles Hamiltonian is defined by $H=\frac{1}{2}(\dot{x}^2+\dot{y}^2)+\frac{1}{2}(x^2+y^2)+x^2y-\frac{1}{3}y^3$ and is a well-known model used in galactic dynamics among other contexts. In addition, it constitutes a paradigm in Hamiltonian nonlinear dynamics as a two-dimensional time-independent dynamical system. One relevant feature of this system is that for energy values below a critical value $(E= 1/6)$ orbits are bounded, and above this critical energy orbits are unbounded. As a matter of fact,  for energies above the escape energy $(E= 1/6)$, initial conditions launched from the center have three different exits to escape from the potential well ultimately to the infinity. The different colors of the basins, i.e., red, green and blue, represent the initial conditions that eventually leave the potential well through one of the three different exits in physical space. We can observe that basins in panels (a) and (b) have the same fraction ($1/3$) of the area of the phase space due to its $2\pi/3$ symmetry. However, one would intuitively attribute a higher unpredictability to the figure displayed in (a) than in (b). Put it another way: if chosen at random, it would be easier to predict initial conditions in (a) than in (b). This difference can be measured by the the uncertainty dimension $\alpha_a<\alpha_b$. However, panels (b) and (c) have exactly the same boundaries due to the merging property of the Wada basins~\cite{daza2018ascertaining}. In fact, picture (c) has been constructed replacing the red and blue pixels by yellow pixels. We can imagine that the upper and left exits are somehow connected and therefore there is only one basin associated to both of them. So we have that $\alpha_b=\alpha_c$, although in this case their area would be different, since the yellow basin occupies $\frac{2}{3}$ of the total area. Therefore, we can see how these ingredients grasp some aspects associated to the unpredictability of the basins, but it is easy to find cases where they fail to correctly quantify it. This situation has given rise to offer vague affirmations when studying basins. The basin entropy appeared as a natural way to answer the issue of the unpredictability associated to the basins, integrating some of the pre-existing concepts and providing a conceptual framework to classify the basins.

\section{What is the basin entropy?}
In multistable dynamical systems, the asymptotic behavior is uniquely defined by the initial conditions. We assume the function $y=B(x)$ to be known, where $y\in[1,N_A]$ labels the different asymptotic behaviors and $x$ accounts for the initial conditions. Although we can define such relation mathematically in a deterministic way, in practice the initial conditions $x$ typically have some uncertainty associated. This means that in order to study the uncertainty of a multistable system, initial conditions should not be regarded as points in some state space, but as balls with some uncertainty radius $\varepsilon$. Some of these balls fall entirely within a basin, meaning that despite the initial uncertainty there is no doubt concerning their fate. However, others will fall in the boundaries, meaning that their final state would be unknown. The idea of the basin entropy is precisely to quantify those uncertain states. For that purpose, we compute the probabilities $p_{i,j}$ associated to the different basins $j=1,\ldots,N_A$ for each $\varepsilon$-ball $i$. If we use a large enough number of balls $N$, we can define the basin entropy as the average value of the entropy of all the balls,
\begin{equation}
S_b=-\frac{1}{N}\sum\limits_{i=1}^{N} \sum\limits_{j=1}^{m_i} p_{i,j} \log \left( p_{i,j}\right). \label{eq:entropy_definition}
\end{equation}
Thus, the basin entropy can take values $S_b\in[0,\log N_A]$. With a few extra assumptions, we can get some insights about the meaning of the basin entropy. If we assume that the probabilities of the different fates are equally distributed within each ball $p_i=1/N_A$ $\forall i$, and that we only have one boundary, then we get
\begin{equation}
S_b= \frac{n_k}{\tilde n} \varepsilon^{\alpha} \log (N_A), \label{eq:3terms}
\end{equation}
where $\alpha$ is the uncertainty exponent, defined as the difference between the topological dimension of the basins and the fractal dimension of the boundary $\alpha=D-d$. This expression allows us to identify three key ingredients of the basin entropy: the lacunarity (associated to the first constant $\frac{n_k}{\tilde n}$), the uncertainty exponent and the number of attractors. 

If instead of considering all the $N$ balls, we just take into account the $N_b$ balls ($N_b\leq N$) falling on the boundary, we can define the new concept of boundary basin entropy
\begin{equation}
S_{bb}=-\frac{1}{N_b}\sum\limits_{i=1}^{N_b} \sum\limits_{j=1}^{m_i} p_{i,j} \log \left( p_{i,j}\right).
\label{eq:Sbb}
\end{equation}
This quantity is most convenient and can be used to study the fractality of the boundaries. In particular, the $\log 2$ criterion and other improved statistical tests~\cite{puy2021test} can be used as a sufficient (but not necessary) condition for fractality without using different magnification scales.

\section{Multidisciplinary applications of the basin entropy}
As already commented earlier, the basin entropy has been applied to a wide variety of scientific fields. In many problems of astrophysics, such as the restricted n-body problems, it is required to quantify the associated unpredictability and the basin entropy provides the ideal tool~\cite{zotos2020basins, zotos2018basins, zotos2018basinsb, dubeibe2018dynamical, zotos2018basinsa}. Furthermore, in the paradigmatic Hénon-Heiles system, the purpose of the basin entropy has gone beyond measuring the unpredictability associated to its basins, and it has been also exploited as a tool to quantify the changes in the KAM islands of the nonhyperbolic regime~\cite{nieto2020measuring}. These changes alter the escape time distribution and can be easily identified by measuring the basin entropy, while inadvertent by the fractal dimension.

The basin entropy has been proved fruitful in other fields in physics too. For example, in~\cite{daza2017chaotic} the boundary basin entropy is proposed as a method to detect fractal structures in experiments with cold atoms, delving deeper in the correspondence between classical and quantum chaos.  Another example in the world of the tiny is the use of the basin entropy to analyze suspended beam micro/nanoelectromechanical (MEMS/NEMS) resonators actuated by two-sided electrodes~\cite{gusso2019nonlinear}. The complex nonlinear behavior and the multistability of such system was carried out by using basins of attraction. The results for the uncertainty exponents were compared with those obtained using the basin entropy method, concluding that the basin entropy provides a reliable alternative method of calculation. Another application refers to plasma dynamics~\cite{mathias2017fractal}, where the basin entropy has provided a useful framework to give a quantitative description of the basin structures. Both basin and basin boundary entropy were found to depend on the perturbation strength as it does the set of initial conditions leading to an escape through the tokamak wall. Also in plasma physics, we can mention transport problems in chaotic area-preserving nontwist maps~\cite{mugnaine2018dynamical}. The basin entropy study was combined with the numerical computation of the transmissivity across the internal transport barrier, allowing to identify the basin boundary of the escape regions related to transport across the barrier. In a biophysical model ~\cite{silva2020transport}, the basin entropy has been particularly convenient to investigate the advection of blood particles in the carotid bifurcation in a healthy scenario, due to the relevant role played not only by the area of the basin but also by the basin boundary topology. 

In a problem in condensed matter physics~\cite{cisternas2021stable}, a small number of XY magnetic dipoles subject to an external magnetic field was used for studying the origin of their collective magnetic response.  Basins of attraction for the dipoles were constructed, and the boundary basin entropy facilitated the analysis of the complexity of the solutions, finding that the damping timescale is critical for the emergence of fractal structures.  Besides those applications already mentioned, the idea of basin entropy has been also applied to account for special relativistic effects on celestial dynamics problems~\cite{bernal2018uncertainty, bernal2020influence}.  Furthermore,  in a general relativity context, the calculation of the basin entropy showed that the uncertainty in predicting the final exit state increases with stronger magnetic interactions in a weakly magnetized Schwarzschild black hole~\cite{bautista2021chaotic}. Another original application of the basin entropy appeared in ~\cite{mugnaine2019basin}, in the context of biodiversity for the analysis of a stochastic network evolution of a cyclic three-species system. As a matter of fact, the main idea is that the basin entropy concept might be useful to be applied to any physical system showing multistability. 

\begin{figure}
%\onefigure{Sbb.pdf}
\includegraphics[width=0.4\textwidth]{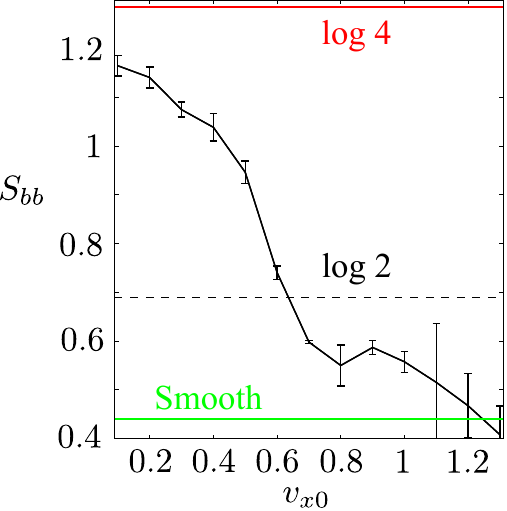}
\caption{The basin boundary entropy $S_{bb}$ versus the launching velocity $v_{x0}$ of cold atoms in a double beam setup~\cite{daza2017chaotic}. The horizontal lines show the maximum possible value for the four attractors $\log 4$, the original $\log 2$ criterion for fractality~\cite{daza2016basin}, and the test developed by Puy et al.~\cite{puy2021test}. These methods allow to test the presence of fractal structures at a given resolution, which is particularly important in hard to escalate experiments like this.}
\label{fig:Sbb}
\end{figure}

Definitely, the basin entropy has been widely applied to problems in nonlinear dynamics and fractal geometry. Gusso et. al. \cite{GUSSO2021111532} proposed another unexpected application of the basin entropy. They used it as a way to compute the fractal dimension of the basins of attraction. With an approach inspired from the box-counting algorithm, the authors achieved a high level of accuracy for two dimensional basins. 

The Wada property was one of the initial motivations for the investigation of the basin entropy and both of them have strong connections. In this special case of fractal structures, three or more basins share a common fractal boundary. In the literature, Wada basins were regarded as even more unpredictable than fractal basins. In the framework of the basin entropy we can understand the particularities of the Wada property: it maximizes the uncertainty associated with boundaries, but not with basins. Different parameters have been defined to characterize the Wada property. Among them, the Wada index is closely related to the basin entropy~\cite{saunoriene2021wada}. In fact, for basins with the property of Wada, it can be verified that
\begin{equation}
   S_{bb}=-\frac{log N_A}{N_A} W, 
 \label{eq:wSbb}
\end{equation}
where $S_{bb}$ is the basin boundary entropy, $N_A$ is the number of attractors and $W$ is the average of the Wada index over the boundary.

\section{Fractals at a single scale}

Fractals are mathematically defined using infinite magnification scales, but there are many practical situations where one would like to detect fractality using only one single scale. If we think in terms of the boundary basin entropy $S_{bb}$, a smooth boundary typically separates two possible fates and therefore it contains one bit of information. However, fractal boundaries wander in complex meanders which need more information to be described. The $\log 2$ criterion is a sufficient (but not necessary) condition for fractal boundaries that exploits this idea. It was first introduced in the original paper of the basin entropy and the reasoning was quite simple. In smooth basins, boundaries separate three basins in a few countable number of cases. Therefore, if the $S_{bb}$ value corresponds to a situation where we have mostly boundaries separating more than two basins, those boundaries must be fractal. Indeed, it is possible to show that
\begin{equation}
   S_{bb}>\log 2 \Rightarrow \alpha<1, 
\label{eq:log2}
\end{equation}
where $\alpha$ is the uncertainty exponent. This means that if the $S_{bb}$ is computed and is larger than $\log 2$, we have strong evidence of the presence of fractal boundaries. The converse is not true and indeed the $\log 2$ criterion does not work for bistable systems. The main advantage with respect to other methods to determine whether a boundary is fractal or not, is that it can be applied using a single scale. Therefore this method is particularly well suited for experimental setups where it is often the case that the scale cannot be changed at will. Nonetheless, the $\log 2$ criterion could be further improved. Following the goal of a fractal detection within one single scale, a new statistical method was recently devised based on the basin entropy concept~\cite{puy2021test}. The idea is that one can study the statistical properties of $S_{bb}$ for smooth boundaries as a way to identify them. Significant deviations from such values suppose important indicators of non-smoothness, that is, of fractality. It can be shown that if the $S_{bb}$ is computed using disks in two dimensions, then we have
\begin{equation}
   S_{bb}\neq 0.4395093(6)\pm \sigma \Rightarrow \alpha<1, 
 \label{eq:0.43}
\end{equation}
where $\sigma$ is the error that typically appears when the basin entropy is computed and that must be estimated for each case. Equation~\ref{eq:0.43} is clearly more restrictive than Eq.~\ref{eq:log2}, and consequently it allows a more precise detection of fractal boundaries. We can see a comparison of these criteria in Fig.~\ref{fig:Sbb}. The black solid line is an estimation of the $S_{bb}$ value for a system with four basins as the parameter $v_{x0}$ is varied. In this case, $S_{bb}=log 4$ is the maximum possible value, which would imply completely fractalized basins. As $v_{x0}$ decreases, the $S_{bb}$ approaches this asymptotic value. In the analyzed range under consideration, we can see that only values with $v_{x0} \lesssim 0.6$ fulfill the $log 2$ criterion. However, the statistical method (green line) shows that we have fractal basins almost for the entire range of values. This evinces how the statistical test is more precise than the $\log 2$ criterion and can be used to detect fractal basins using only one scale.

\begin{figure*}[!ht]
\begin{center}
\includegraphics[width=\textwidth]{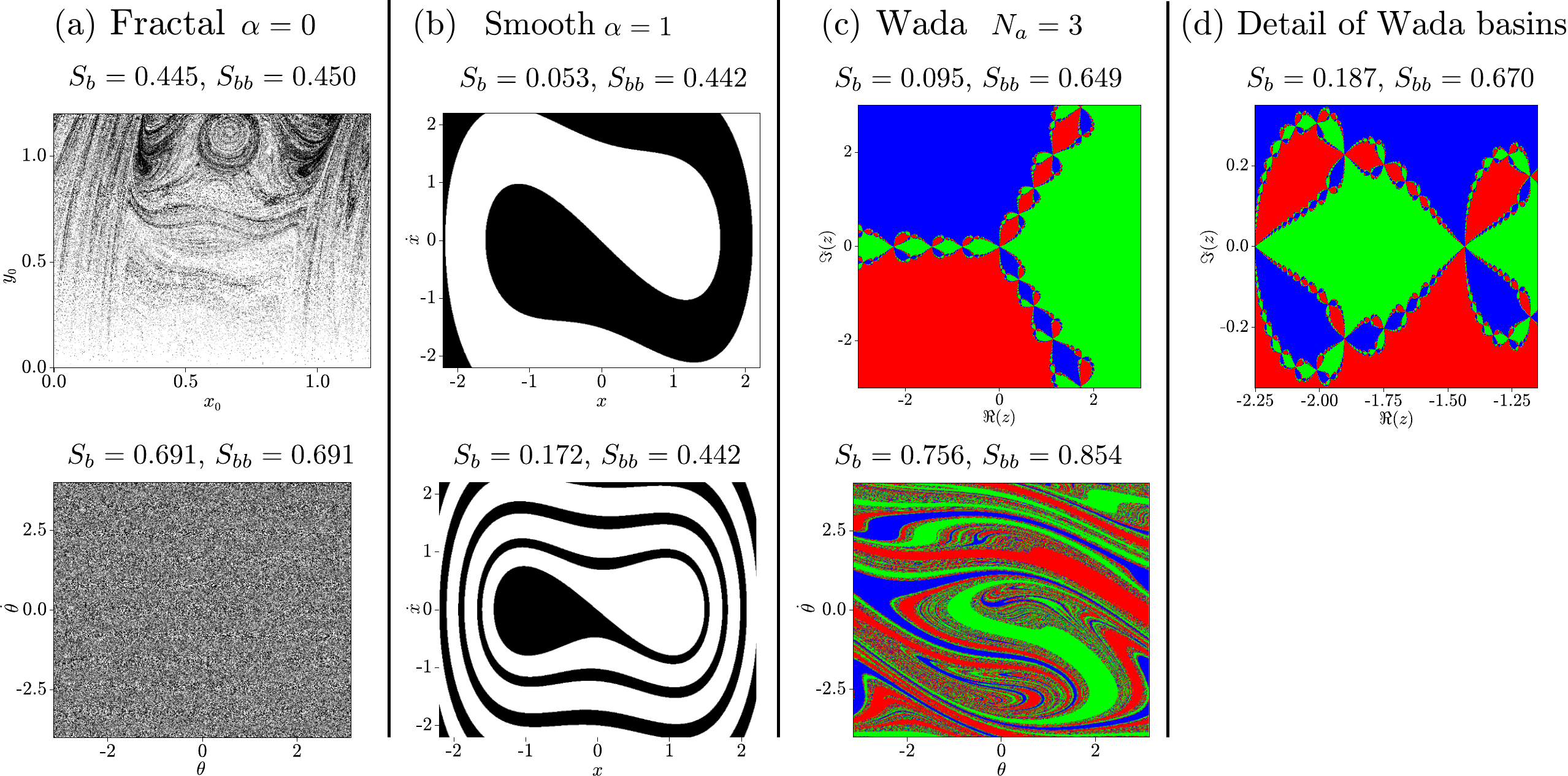}
\end{center}
\caption{{\bf Basins classified according to the basin entropy.} In (a) we represent two riddled basins obtained from the dynamical system published in \cite{ott1994transition} on the left and the forced damped pendulum $\ddot \theta + 0.2 \dot \theta + \sin \theta = 1.3636 \sin 0.5 t$ on the right. In (b) the two basins with a smooth boundary have been obtained from the periodically driven Duffing oscillator $\ddot{x} + d \dot{x} - x + x^3 = 0.1 \sin 0.1 t$ with $d= 0.4$ on the left and $d=0.1$ on the right. (c) Represents two examples of Wada basins with three attractors. On the left, the basins for the Newton algorithm to find the roots of the polynomial $f(z) = z^3 -1$ in the complex plane. The figure on the right shows the Wada basins of the forced damped pendulum $\ddot \theta + 0.2 \dot \theta + \sin \theta = 1.66 \sin t$. (d) The figure represents a portion of the phase space of the Newton fractal.  The basin entropy has been computed on a grid of 1000$\times$1000 initial conditions. All figures have been computed with Julia programming language using an automatic algorithm \cite{datseris2022}. The code for reproducing the plots is available at \cite{github_code}.}
\label{fig:class}
\end{figure*}

\section{Basin entropy as a classification tool}

The previous discussion lays out a framework for the classification of the basins \cite{DAZA2022Class} using as a criterion the basin entropy. Equation~\ref{eq:3terms} possesses three terms that measure different aspects of basins of attraction. The first term, the lacunarity, quantifies the {\it gapiness} of the structures present in the basins. An extreme case of a boundary with high lacunarity is a basin in the form of a set with infinite initial conditions with a zero set measure known as Cantor dust. The second term measures the final state sensitivity of the basins with uncertainty exponent $\alpha$. The value $\alpha = 0$ corresponds to a complete fractalization of the phase space known as riddled basins, whereas on the other end smooth boundaries will have an exponent $\alpha = 1$. All other basins take a value in the interval $0 < \alpha < 1$. The third and last term accounts for the number of possible attractors $N_a$ in the basins. 

The basin entropy takes characteristic values for basins that maximize one of the three ingredients of Eq.~\ref{eq:3terms}. This classification process is illustrated in Fig.~\ref{fig:class} with six different basins. The panel (a) represents two riddled basins with the same uncertainty exponent $\alpha = 0$. Despite the extraordinary complex aspect of both pictures, it is clear that their fine structure is different. Since $\alpha$ and $N_a$ are the same for the two figures, the values of the basin entropy reveal the difference of lacunarity between the two basins. For riddled basins the condition $S_b \simeq S_{bb}$ is fulfilled but the bottom panel has the value $\log 2$, the highest possible value for basins with two attractors. The basin entropy allows to clearly establish the order in uncertainty between the two examples. Panels (b) of Fig.~\ref{fig:class} depict two examples of basins with a smooth boundary, meaning that the uncertainty exponent takes the maximum value $\alpha = 1$. Both boundaries have a similar structure, but the basin entropy sorts the two plots according to the intricacy of their boundary. Notice that the value of the boundary basin entropy is close to $0.439$,  which is the theoretical value of a smooth boundary. Still the intuition automatically assigns a larger unpredictability to the basins on the bottom plot and the basin entropy confirms this first thought. Figure~\ref{fig:class} (c) represents two basins with the Wada property. Using the Wada index and the basin entropy it is possible to identify the Wada property for this system. Nevertheless, a simple look at the two basins is enough to decide which basins are the most intricate. The basin entropy quantifies this contrast with two very different values. It turns out that basins with the Wada property are not necessarily more intricate than other basins. It is a peculiar attribute of the boundary but other aspects of the basins contribute to its complexity too. 

The direct classification between two basins is possible when both basins maximize one of the ingredients of the basin entropy. But the comparison between two basins that have different characteristics also bears its fruits. The basins with a smooth boundary in Fig.~\ref{fig:class} (b)-bottom have a basin entropy higher than the fractal basins in Fig.~\ref{fig:class} (c)-top. This counterintuitive fact is an example of how difficult it is to establish a hierarchy between basins observed with a finite resolution. The area of the phase space occupied by the boundary of the Newton fractal is smaller than the smooth boundary of the basins with two attractors of the Duffing oscillator. However, if we zoom in on the Newton fractal to enlarge a part of its phase space as seen in Fig.~\ref{fig:class} (d), the basin entropy is now $S_b = 0.187$: a value above the basins with the smooth boundary. It can be interpreted as a limitation of the basin entropy, while it is in fact a problem of interpretation of the unpredictability of the basins. The question ``Which figure is the most intricate?'' has not always a clear cut answer when we deal with basins at a finite resolution. The basin entropy gives an answer according to the blending of three ingredients that cannot always be decomposed unequivocally. 

\section{Future perspectives}
As we have described earlier, the basin entropy has become a fundamental tool for understanding unpredictability in multistable systems. Nevertheless, we believe that it has a very promising potential for its use in the future in numerous research problems. Among the conceivable applications, we can think of the basin entropy as a way to predict and classify bifurcations in dynamical systems. In fact, when a new attractor and its basin appear or disappear as a parameter is modified, there is a sudden change in the basin entropy value. A connection between bifurcations and basin entropy values could be an alternative for the study of the parameter dependence of dynamical systems. It might be even feasible to detect some early indicators of bifurcations that are not easily observed with current methods. The relationship of the basin entropy with other quantities such as the uncertainty exponent or the Wada index has shed light into different areas of nonlinear dynamics. This kind of connections could be further extended to other quantities and contexts. Maybe the Kaplan-Yorke conjecture can provide such a link between the Kolmogorov-Sinai entropy and the basin entropy, creating a bridge between instantaneous and asymptotic dynamics. Stochastic dynamics have also been studied in combination with basins of attraction~\cite{nieto2021final, serdukova2016stochastic}. In these cases, the basin entropy could also be adapted to provide information about the system final state unpredictability.

Another natural development of the basin entropy is related to the Wada property. Actually, the detection and characterization of Wada basins has drawn much attention in recent years~\cite{wagemakers2020detect}. In the same vein, we can imagine a new way to detect Wada boundaries based on the boundary basin entropy $S_{bb}$. The $\log 2$ and then the statistical criterion allowed us to identify fractal basins at a given scale. A similar reasoning could lead to a sufficient condition to detect Wada boundaries at a single scale, which would be quite advantageous for some experiments. Although some of the applications of the basin entropy are related to the use of just one scale, one could aim to define a free-scale related quantity, by properly integrating the information provided by the basin entropy at different scales in some sort of renormalization process. For example, slim fractals with varying properties along different scales typically appear in undriven dissipative systems~\cite{chen2017slim}. 

The field of complex networks can also benefit from the basin entropy ideas. Recently, Halekotte et al. ~\cite{halekotte2021transient} have studied the multistability of the British Power grid with the basin entropy. It would be interesting to investigate the occurrence of fractal boundaries in such high dimensional systems. The statistical criterion of Eq.~\ref{eq:0.43} can be extended to n-dimensional basins~\cite{puy2021test}. Therefore, it would provide a useful test for high-dimensional systems such as complex networks.

Nonetheless, as usually happens in research, probably some of the most exciting developments concerning the basin entropy do not figure in the previous lines, but still remain to be unveiled. The research on multistable systems and their unpredictability will definitely expand in physics and other scientific disciplines in the upcoming years.

\section{Conclusions}
The basin entropy constitutes a new tool to measure the final state unpredictability of dynamical systems by analyzing their basins. Different domains in Physics, such as cold atoms, shadows of binary black holes, and classical and relativistic chaotic scattering in astrophysics have been benefited so far of its use. We believe that the idea of basin entropy will become a relevant tool in the study of complex systems with applications in multifarious scientific fields. Many disciplines in science and engineering have recently received a tremendous influence from nonlinear dynamics. We would like to emphasize that the field of chaos theory can provide useful tools to understand the rich dynamics of many fundamental problems in physics, something that could be achieved through fruitful scientific interactions.

\acknowledgments
The authors acknowledge financial support from the Spanish State Research Agency (AEI) and the European Regional Development Fund (ERDF, EU) under project PID2019-105554GB-I00.

\bibliography{biblio_EPL}

\bibliographystyle{eplbib}

\end{document}